\documentclass[aps,preprint]{revtex4}%
\usepackage{amsfonts}%
\usepackage{amsmath}%

\usepackage{graphicx}
\def\be#1{\begin{equation}\label{#1}}
\def\ee{\end{equation}}
\def\bea#1{\begin{eqnarray}\label{#1}}
\def\eea{\end{eqnarray}}
\def\sp{\hspace{.5em}}
\def\Eq#1{Eq.(\ref{#1})}
\def\Fig#1{Fig.(\ref{#1})}

\def\no{\nonumber \\}

\def\mbf#1{\mbox{{\boldmath $#1$}}}

\def\expm2piOmega{e^{-2\pi\Omega}}

\def\sp{\hspace{.25em}}

\begin{document}
\title{The Unruh effect in an Ion Trap: An Analogy}
\author{Paul M. Alsing}\email{alsing@hpc.unm.edu}
\affiliation{Department of Physics and Astronomy, 800 Yale Blvd NE, University of New Mexico, Albuquerque, NM 87131}
\author{Jonathan P. Dowling}\email{jdowling@baton.phys.lsu.edu}
\affiliation{Department of Physics, Louisiana Sate University, Baton Rouge, Louisiana }
\author{G.J. Milburn}\email{milburn@physics.uq.edu.au}
\affiliation{The University of Queensland, Centre for Quantum
Computer Technology, Department of Physics,School of Physical Science, Brisbane, Australia, }

\begin{abstract}
We propose an experiment in which the phonon excitation of ion(s)
in a trap, with a trap frequency exponentially modulated at rate
$\kappa$,  exhibits a thermal spectrum with an "Unruh" temperature
given by $T=\hbar\kappa$. We discuss the similarities of this
experiment to the usual Unruh effect for quantum fields and
uniformly accelerated detectors. We demonstrate a new Unruh effect
for detectors that respond to anti-normally ordered moments using
the ion's first blue sideband transition.
\end{abstract}

\date{\today}
\maketitle

\section{Introduction}
It has been known for many decades that an accelerated detector, moving in a quantum field prepared in the ground state
of the field modes for an inertial frame, will become excited\cite{Davies,Unruh, BD}. In the case of constant acceleration,
$a$,  the frequency response of the detector is completely equivalent to the response of an inertial detector in a thermally
excited field with a temperature $T$ given by $T=\hbar a/2\pi c$. We call this Unruh-Davies radiation.  Quite clearly, such
an effect would be very difficult to see given current technologies. In this paper we suggest an analogous system, based on
detecting phonons of the vibrational modes of cold trapped ions\cite{LeibfriedA,scully}.  In many ways this parallels a
theme, pioneered by Unruh, of  sonic equivalents for quantum fields in curved space time\cite{Unruh-sonic}.

Our analogy is based on an alternative view of Unruh-Davies radiation in terms of the time dependent
red shift seen by an accelerated observer\cite{alsing_milonni}. By controlling the trapping potentials
of trapped ions it is possible to modulate the normal mode frequencies so that they have the same time
dependent phase as red-shifted frequencies seen by a constantly accelerated observer. Suppose now that
the ions are prepared (using laser cooling) in the ground state of the normal modes of vibration of the
time independent trap.  If a suitable detector of the vibrational quanta for the trapped ions  was available,
they could be used to detect excitations out of the ground states of vibrational motion due to the frequency
modulation.  This would be analogous to the response of  an accelerated detector to a scalar field prepared
in a Minkowski ground state. Fortunately the ions themselves can be made to respond as phonon detectors.

In ion trap implementations of quantum information processing, a laser is used to couple the vibrational
motion of a trapped ion  to an electronic transition between states which we denote $g\leftrightarrow e$ \cite{DJ}.
Furthermore it is possible to readout one or the other of these internal electronic states, say $g$,  using a
laser (the readout laser) to drive a cycling transition between $g$ and another electronic state, thereby
producing fluorescence conditional on whether the ion was in state $g$.  Such measurements are highly
efficient and closely approximated by a perfect projective measurement of the internal electronic state.
In effect this scenario defines a  {\em phonon detector} that may be turned on and off at will.  To be more
specific, we can implement various kinds of phonon detectors by carefully tuning the laser frequency $\omega_L$
to one of the vibrational sidebands of the ion.  This enables one to realize rather unconventional phonon
detectors that respond to antinormally ordered moments (blue sideband) as well as the more conventional
normally ordered moments (red sideband) as we explain in more detail below.

\section{Trapped ion model.}

The interaction Hamiltonian describing the coupling of the internal and vibrational degrees of freedom of the $m$'th ion
in a linear array of ions in a trap (in the interaction picture, and Lamb-Dicke limit) can be written as \cite{DJ}
\be{1}
\hat{H}^{(m)}_I = \hbar\Omega_0 k \cos\theta \hat{q}_m(t)\sigma_x(t)
\ee
where
\be{sigmax}
 \sigma_x(t)=e^{-i\Delta t} \sigma_- + \textrm{h.c.},
\ee
and $\Omega_0$ is the (scaled) Rabi frequency, $\Delta \equiv \omega_A - \omega_L$ is the detuning
between atomic resonance and the laser, $k=\omega_L/c$ is the wavevector,  $\theta$ is the angle the laser beam
makes with the longitudinal axis for the linear ion chain, $\hat{q}_m(t)$ is the quantized local displacement
of the $m$th ion about its equilibrium position and $\sigma_-= | g \rangle \langle e|$ is the atomic
lowering operator between the upper and lower atomic states $| e \rangle$ and $| g \rangle$ separated
by frequency $\omega_A$. We assume the equilibrium position of the $m$th ion is located at a node of
the standing wave laser beam, we have used the rotating wave approximation to describe the interaction between the
laser and the ion. In the Lamb-Dicke limit, terms of order $\hat{q}^2_m(t)$
have been neglected since the displacement of the ion is much less than
the wavelength of light. In an experiment the $e\leftrightarrow g$ is a quadrupole transition and is driven by a Raman process.
Thus $\Omega_0$ is an effective Rabi frequency for this process.

Equation (\ref{1}) could also be regarded as a discretised representation for the interaction of a scalar field,
$\hat{\phi}(x)$, and a local detector, with transition frequency $\Delta$, where $\hat{\phi}(x=ml)=\hat{q}_m$ for
some discretisation length $l$.  In such an interpretation the interaction Eq.(\ref{1} is equivalent to the
Unruh model of a particle detector\cite{Unruh,BD}, with only two internal energy levels.

Note that  the detector  can be turned on and off through the dependance on the external laser field in $\Omega_0$,
a somewhat unusual feature for  field quanta  detectors.  Another unusual feature of this detector is that the transition
 frequency of the detector, $\Delta=\omega_A-\omega_L$ can be varied by tuning the external laser. Conventional detectors
 would have a fixed transition frequency. This latter feature will enable us to define different kinds of phonon detectors.

The local displacement of the $m$th ion, $\hat{q}_m(t)$, can be expanded  in
terms of creation and annihilation operators for \textit{global} normal modes (phonons) of the $N$-ion system
$\hat{Q}_p $ by \cite{DJ}
\bea{2}
\hat{q}_m(t) &=& \sum_{p=1}^N b^{(p)}_m \hat{Q}_p(t) \no
& = & \imath \sqrt{\frac{\hbar}{2 m \nu N}} \sum_{p=1}^N s^{(p)}_m
\left( \hat{a}_p e^{-i\nu_p t} - \hat{a}_p^\dagger e^{i\nu_p t}\right)
\eea
where the coupling constant is defined by
\be{3}
s^{(p)}_m = \frac{\sqrt{N}b^{(p)}_m}{\mu_p^{1/4}}.
\ee
In the above $\nu_p$ are the normal mode trap frequencies given by $\nu_p = \sqrt{\mu_p} \nu$ in
terms of the bare trap frequency $\nu$ and the eigenvalues $\mu_p$.  For the center of mass
mode $s_m^{(1)}=1$, $\nu_1 = \nu$ and for the breathing mode $\nu_2 = \sqrt{3} \,\nu$,
$s_m^{(2)}= \sqrt{N} \, (\sum_{m=1}^N \, u_m^2)^{-1/2} \, u_m\,/\;\sqrt[4]{3}$, where in
the later $u_m$ are the components of the breathing mode eigenvector $b^{(2)}_m = u_m/\sqrt{\sum_{m=1}^N u_m^2}$
normalized to unity.
Further normal mode parameters for up to $N=10$ ions are computed numerically in James \cite{DJ}.

Defining the Lamb-Dicke parameter as $\eta = \sqrt{\hbar k^2\cos^2\theta/2m\nu}$ we can write
our Hamiltonian in its final form
\be{5}
\hat{H}^{(m)}_I  = \chi\hat{q}_m(t)\sigma_x(t)
\ee
where we have defined the operator
\begin{equation}
\hat{q}_m(t)=\frac{i}{\sqrt{N}} \sum_{p=1}^N s^{(p)}_m
\left( \hat{a}_p e^{-i\nu_p t} - \hat{a}_p^\dagger e^{i\nu_p t}\right)\ ,
\label{field}
\end{equation}
and $\chi=\Omega_0\eta$.
This Hamiltonian describes the coupling between the two level electronic transition (the detector) and the vibrational
degrees of freedom whenever the external laser is turned on ($\Omega_0\neq 0$). Note that we do not make any
assumptions at this stage about the relative size of $\Delta$ and $\nu_p$. We wish to keep $\Delta$ as a free
parameter which may be varied to define different kinds of phonon detectors. In analogy with the Unruh detector
model, the field operator $\hat{q}_m(t)$ represents the scalar field at the position $x=ml$ and
time $t$,  $\hat{q}_m(t)=\hat{\phi}(x,t)$. In physical terms this is the displacement of the $m$th ion as a function of time.

It is worth noting an important difference between this model and the usual treatment of a particle detector.
In the case of a usual detector  the frequency term, $\Delta$, would be strictly positive thus defining the
positive and negative frequency components of the dipole $\sigma_\pm$.  In \Eq{sigmax}, the parameter $\Delta$
can be positive or negative so we cannot simply refer to positive or negative frequency components in absolute terms.
However the operators $\sigma_\pm$ will retain their usual definition as raising and lowering operators.

\begin{figure}[h]
\includegraphics[width=4in,height=3in]{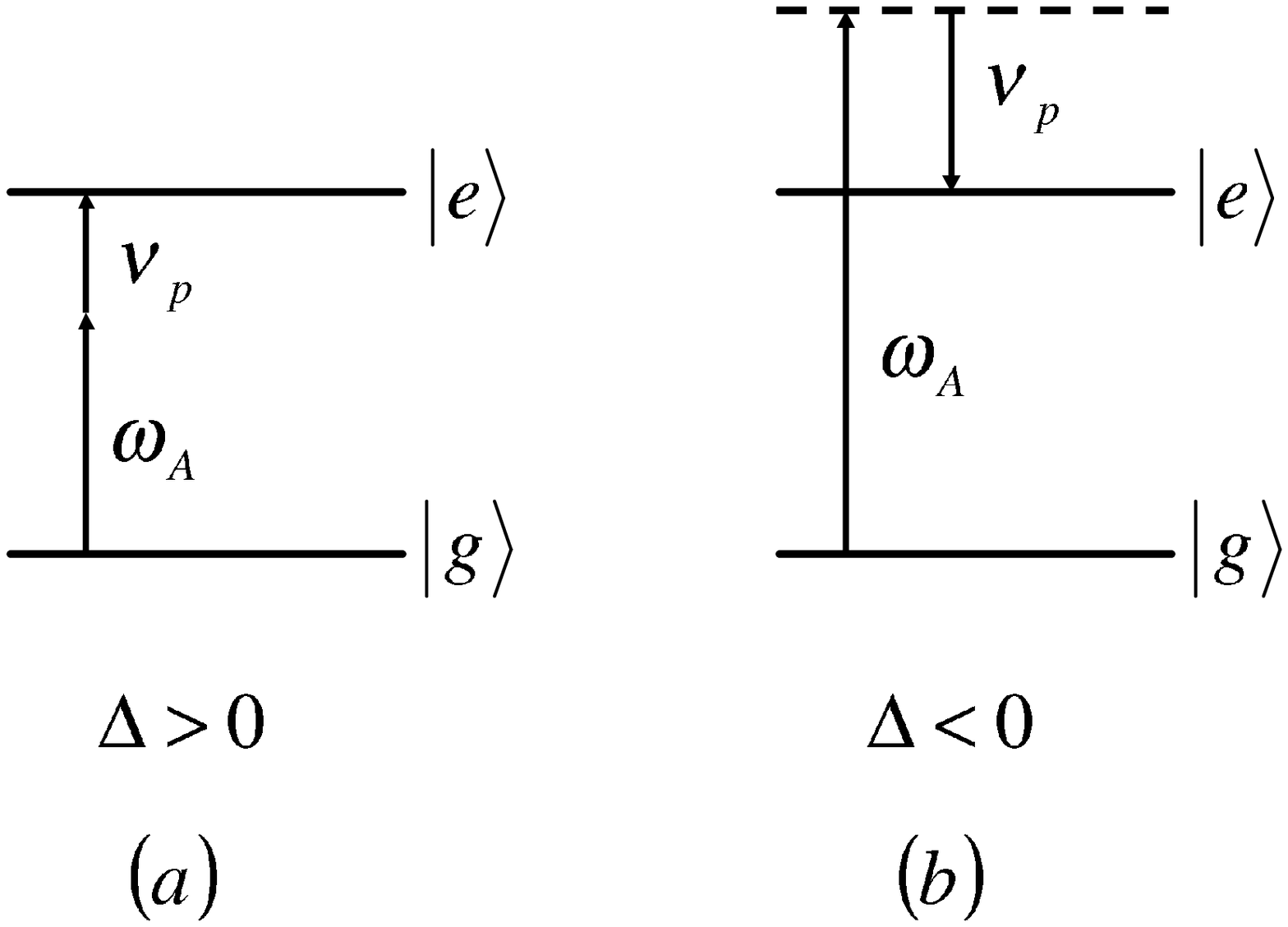}
\caption{Energy Level Diagrams. Tuning to (a) red side band, $\Delta \equiv\omega_A - \omega_L = \nu_p > 0$, (b)
blue side band $\Delta = -\nu_p< 0$. The Unruh effect \Eq{14} takes place on the red sideband. We obtain an
\textit{anti-normally ordered} Unruh effect if we tune to the blue side band. }
\label{fig1}
\end{figure}

 In the case that $\Delta>0$, the laser is detuned below the atomic transition, which we refer to as red detuning.
 We can resonantly excite so called red sidband transitions when $\Delta\approx n\nu_p$. Near such a resonance ($n=1$)
 we can make the rotating wave approximation and describe the interaction by the Hamiltonian
 \begin{equation}
 \hat{H}^{(m)}_I = \imath \frac{\Omega_0\eta}{\sqrt{N}} \sum_{p=1}^N s^{(p)}_m
\left( \hat{a}_p \sigma_+ - \hat{a}_p^\dagger \sigma_-\right )
\label{red}
\end{equation}
This describes the usual Jaynes-Cummings model of a two level system interacting with a bosonic degree of freedom.
In physical terms it describes a Raman process in which one laser photon and one trap phonon are absorbed to
excite the atom (see figure \ref{fig1}).  A phonon detector defined this way would respond to the normally
ordered moments of the phonon field amplitude.

In the case that $\Delta <0$, the laser is detuned above the atomic transition, which we will refer to as blue detuning.
The resonant term for the fisrt blue sideband is then given by
 \begin{equation}
 \hat{H}^{(m)}_I = \imath \frac{\Omega_0\eta}{\sqrt{N}} \sum_{p=1}^N s^{(p)}_m
\left( \hat{a}_p \sigma_- - \hat{a}_p^\dagger \sigma_+\right )
\label{blue}
\end{equation}
Again this is a Raman process in which the atom is excited by the absorption of one laser photon and the
{\em emission} of one phonon (see figure \ref{fig1}).  Considered as a phonon counter this would  correspond
to a detector that responded to anti-normally ordered moments of the phonon field amplitude.

Using laser cooling techniques it is possible to cool the system very nearly to the ground state of the
vibrational degrees of freedom. In reality this becomes more difficult as the number of ions,  and thus
normal modes increases. However for our purposes even one ion would suffice. In current experiments the cooling is
sufficiently efficient  to reach the ground state with probability 0.999\cite{LeibfriedB}.
We thus assume an initial state of the form,
\be{6}
\psi(0)\rangle\equiv|\mathbf{0}\rangle = |g\rangle\otimes|0\rangle_\textrm{vib}
\ee
where the initial vibrational state is a tensor product of the ground states of each of the normal modes,
\begin{equation}
|0\rangle_\textrm{vib}=\prod_{n=1}^p |0_p\rangle\ .
\end{equation}

We will exponentially chirp the trap frequency up or down such
that \be{7} \nu \to \nu(t) = e^{\pm \kappa t} \ee with $\kappa$
the chirp rate and focus our laser on the first ion ($m=1$). For a
constant trap frequency the phonon annihilation operator satisfies
the usual uncoupled mode equation $\dot{\hat{a}}_p(t) = -\imath
\nu_p \hat{a}_p(t)$ with solution $\hat{a}_p(t) = \exp(-\imath
\nu_p t)\, \hat{a}_p(0)$, $\nu_p = \sqrt{\mu_p}\, \nu$ which was used in \Eq{field}.
For an chirped trap frequency given by \Eq{7}
between an initial time $t_0$ and final time $t$, the phonon mode
now satisfies the following equation of motion, with solution
\be{8} \frac{d}{dt} \hat{a}_p(t) = -\imath \nu_p e^{\kappa t} \,
\hat{a}_p(t)\quad \Rightarrow \quad \hat{a}_p(t) = \exp\left(
\imath\frac{\nu_p}{\kappa} e^{\kappa t_0} \right) \,
               \exp\left(-\imath\frac{\nu_p}{\kappa} e^{\kappa t}   \right)\; \hat{a}_p(t_0).
\ee
where we consider the chirp-up case first (with the chirp down case discussed subsequently).

We now suppose that the coupling to the detector is turned on at the same time as the frequency modulation is turned on,
and turned off at the same time as the frequency  modulation is turned off.  This is a rather different scenario to the
usual discussion of the Unruh-Davies effect in which the detector is always coupled to the field and continuously
accelerated, ie the red shift frequency modulation is always on.

We are interested in the probability $P_m(T,t_0)$ for the excitation of the $m$'th ion from the ground state
to all excited states of detector and field for a detector turned on at $t=t_0$ and turned off at $t=T$.
The detector in our case has only one excited state. We let the excited states of the vibrational degree of
freedom be represented by a complete orthonormal basis $|\phi_\alpha\rangle$.
The total excitation probability is then \cite{note1}
\be{PmG}
P_m(T,t_0)=\chi^2\int_{t_0}^Tdt'\int_{t_0}^T dt'' e^{i\Delta(t'-t'')}G(t',t'')
\ee
where the field correlation function is defined as
\begin{equation}
G(t',t'') = \sp_\textrm{vib}\langle 0|\hat{q}_m(t')\hat{q}_m(t'')|0\rangle_\textrm{vib}
\label{cor}
\end{equation}
with the field now given by
\begin{equation}
\hat{q}_m(t)=\frac{i}{\sqrt{N}} \sum_{p=1}^N s^{(p)}_m \left(
\hat{a}_p e^{-i\frac{\nu_p}{\kappa} e^{\kappa t} } -
\hat{a}_p^\dagger e^{i\frac{\nu_p}{\kappa} e^{\kappa t}}\right)\ ,
\end{equation}
and we have neglected the phase factor arising from the initial
time as it does not contribute to the correlation function.
Substituting this result into Eq.(\ref{cor}),
\be{Pm}
P_m(T,t_0)=\chi^2\sum_{p=1}^N\frac{|b_m^{(p)}|^2}{\sqrt{\mu_p}}|I_p(T,t_0)|^2
\ee
where the integral is
\be{11} I_p(T,t_0) \equiv \int_{t_0}^T
\, dt \, e^{\imath \Delta t} e^{\left(\imath \frac{\nu_p}{\kappa}
\, e^{\kappa t}\right)} .
\ee
With no loss of generality we may
now set $t_0=0$.  We first consider the case $\Delta>0$, which is
the red-sideband case. Our objective is to calculate the
excitation probability for the two level system near the red
sideband transition which we label $P_m^R(T,t_0)$.

To evaluate this integral we first change to dimensionless time,
$\tau=\kappa t$, so that
\be{11a} I_p(T,t_0) =\frac{1}{\kappa}
\int_{0}^{\kappa T} \, d\tau \, e^{\imath a\tau} e^{\imath b \,
e^{\tau}} .
\ee where $a=\Delta/\kappa$ and $b=\nu_p/\kappa$. Next
we define the constant $\alpha$
\begin{equation}
b=e^{-\alpha}
\end{equation}
It then seems sensible to make the new change of variable $\tau'=\tau-\alpha$ so that the integral becomes
\begin{equation}
\frac{e^{ia\alpha}}{\kappa}\int_{-\alpha}^{\kappa T-\alpha} d\tau
e^{ia\tau}e^{ie^\tau}
\end{equation}

In an experiment one needs to vary $\Delta$ near the red or blue
sideband, so we expect that $\Delta$ and $\nu_p$ are of the same
order.  We now consider the limit in which $\nu_p/\kappa <<1$ for
all the normal modes. In this limit $\alpha>>1$, the integral does
not depend much on $\alpha$, so we extend the lower limit of the
integral to infinity. Furthermore we suppose the time over which
the detector and modulation are on is such that $T>> \nu_p$, in
which case the integral does not vary much with $T$. In that case
we can extend the upper limit of the integral to infinity, so that
\begin{equation}
I_p \approx \frac{e^{ia\alpha}}{\kappa}\int_{-\infty}^{\infty}
d\tau e^{ia\tau}e^{ie^\tau}
\end{equation}
Changing the variable of integration to $y=e^{\tau}$ we obtain an expression for $I_p$ with
arbitrary limits as
\bea{Ipy}
I_p &  = & \frac{e^{ia\alpha}}{\kappa}\int_0^\infty dy y^{ia-1}e^{iy} \\
& = &  \frac{e^{ia\alpha}}{\kappa} \Gamma(ia)e^{-\pi a/2}\label{Ip2}
\eea
If we now use the identity $|\Gamma(ix)|^2 = \pi/[x\sinh(\pi x)]$ for real $x$ \cite{GR}
we see that
\be{Ipsqd}
|I_p|^2=\frac{2\pi}{\kappa\Delta}\frac{1}{e^{2\pi\Delta/\kappa}-1}
\ee
Inserting this into \Eq{Pm} we can write $P_m^R$ in the suggestive
form
\be{14}
P_m^R= \frac{\Omega_0^2 \eta^2}{\kappa \Delta} \;
\frac{2\pi}{e^{\hbar\Delta/k_B T}-1} \;
\sum_{p=1}^N \frac{|b_1^{(p)}|^2}{\sqrt{\mu_p}}
\ee
where we have defined the
Unruh temperature as \be{14.1} k_BT\equiv \frac{\hbar\kappa}{2\pi}.
\ee \Eq{14} has the analogous form to a thermal spectrum at
temperature $T$ as seen by a uniformly accelerated observer moving
through a particle-free inertial vacuum. The "thermal" form of the
probability is independent of the phonon frequencies $\nu_p$ since
we have taken the upper limit $t$ to infinity. The last summation
expression is just a numerical factor, which can be computed
\cite{DJ}, or dropped in the case of a single ion in the trap.

The analogy with the Unruh effect \cite{BD} for a uniformly accelerated observer in Minkowski space can be seen as follows.
As discussed in \cite{alsing_milonni} the chirping of the trap frequencies can be considered as arising
from the the modification of the usual Minkowski plane wave $\exp(i\varphi_\pm) \equiv \exp\big(i(k z \pm \omega t)\big)$
due to the motion of the accelerated observer. For an observer moving at constant velocity $v_0$  in Minkowski
space, a Lorentz transformation (LT) of the phase $\varphi_\pm$ viz $c t' = c t \cosh r + z' \sinh r$ and
$z' = z \cosh r + c t' \sinh r$ with the rapidity defined by $\tanh r = v_0/c$, simply transforms
it to $\varphi_\pm \to \varphi'_\pm = (\omega/c) e^{\pm r} \,(z' \pm c t') \equiv (\omega'/c)\,( z' \pm c t')$
which merely produces the constant doppler shifted frequencies in the new frame $\omega' = \omega e^{\pm r}$.
For an accelerated observer, one has to perform a time dependent LT at each instant to the comoving
frame that is instantaneously at rest with respect to the accelerated observer. The orbit of the accelerated (Rindler) observer
as described by an inertial Minkowski observer is given by the Rindler transformation
$c t(\tau) = (c^2/a) \, \sinh (a \tau/c)$ and $z(\tau) =  (c^2/a) \, \cosh (a \tau/c)$ where $a$ is the
constant uniform acceleration. Under this transformation, the Rindler observer experiences phase of the Minkowski
plane wave $\exp(i\varphi_\pm)$ as transformed to
$\varphi_\pm\to \varphi_\pm(\tau) = k z(\tau) \pm \omega t(\tau) = (\omega c/a) \,e^{\pm a \tau/c}$.
These are the chirped frequencies that appear in \Eq{11} in exponentially expanding or contracting the trap.
The Fourier transform of this modified plane wave with respect to \textit{the Rindler proper time} $\tau$
can be considered as a measure of the noise spectrum seen by the accelerated observer viz
\be{14.2}
S(\Delta) =  \left| \int_{-\infty}^\infty \, dt \, e^{\pm\imath \Delta \tau} e^{\left( i \omega c/a \, e^{\pm a\tau/c}\right)} \right|^2
= \left| \displaystyle \frac{(e^{\pm i\pi/2})^{\pm i\Delta c/a}}{(\frac{\omega c}{a})^{\pm i\Delta c/a}}
          \; \Gamma\left( \pm i\frac{\Delta c}{a} \right) \right|^2
= \frac{2\pi c}{\Delta a} \; \displaystyle \frac{1}{e^{i\Delta
c/a}-1}, \ee analogous to $|I_p(\infty,-\infty)|^2$ as in \Eq{Ipsqd}.
Here the perceived thermal temperature is defined from the only
other frequency that can be formed from the Rindler observers
accelerated motion, $a/c$. Thus we get the usual Unruh temperature
defined by $T = \hbar a/ 2\pi c$. In our ion trap analogy, the
role of the acceleration frequency $a/c$ is played by the trap
expansion rate $\kappa$.

\begin{figure}[h]
\includegraphics[width=4in,height=3in]{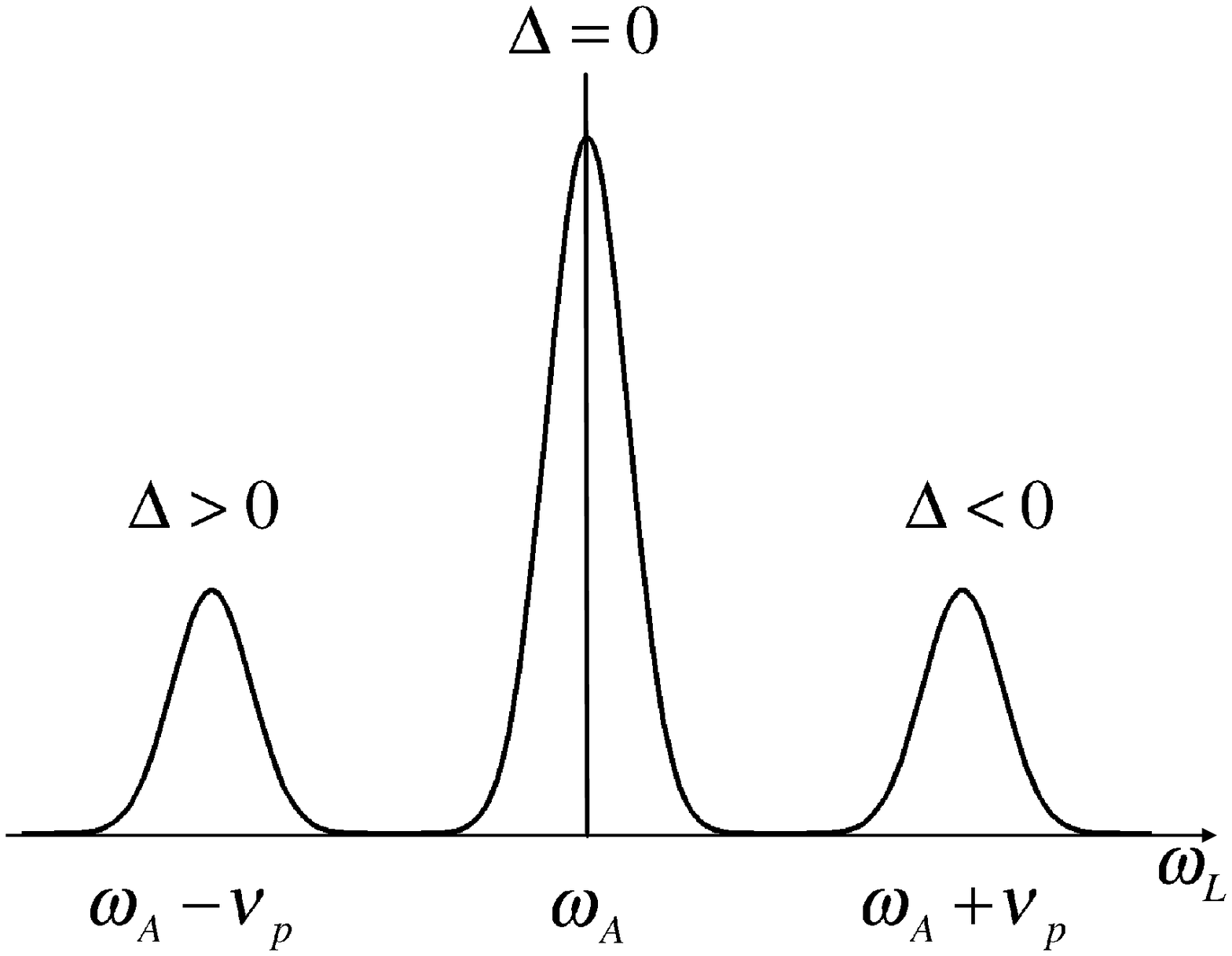}
\caption{Vibrational Resonances. Red (a) and blue (b) vibrational sidebands corresponding to the
energy level diagrams in \Fig{fig1}. }
\label{fig2}
\end{figure}

It is important to note that in order to obtain the Planck factor
$(e^{2\pi\Delta/\kappa}-1)^{-1}$ in \Eq{Ipsqd}, indicative of a
Bose-Einstein (BE) thermal distribution and the signature of the
Unruh effect, we made crucial use of a positive detuning $\Delta >0$,
corresponding to a red sideband detuning in \Fig{fig1} and
\Fig{fig2}. This resulted in the factor $(i)^{i\Delta/\kappa} =
\big(e^{i\pi/2}\big)^{i\Delta/\kappa} = e^{-\pi\Delta/2\kappa}$ that appears in
$I_p$ in \Eq{Ip2}. Dividing the square of this factor into the $\sinh$ function
appearing in the denominator of $|I_p|^2$, resulting from the term $|\Gamma(i\Delta/\kappa)|^2$,
produces the signature BE thermal distribution function.

If on the other hand, we had instead chosen $\Delta = -|\Delta| < 0$
corresponding to a negative detuning to the blue side band \Fig{fig2}{b}, the previous factor would become
$\big(e^{i\pi/2}\big)^{-i|\Delta|/\kappa} =
e^{\pi|\Delta|/2\kappa}$. Dividing the square of this term into the $\sinh$ function
appearing in the denominator then produces alternatively the probability for excitation on the blue sideband
\be{14.5}
P_m^B= \frac{\Omega_0^2 \eta^2}{\kappa |\Delta|} \; \frac{2\pi}{1-e^{-2\pi|\Delta|/\kappa}} \;
\sum_{p=1}^N  \frac{|b_1^{(p)}|^2}{\sqrt{\mu_p}},
\ee
with the same definition of the Unruh temperature as in \Eq{14.1}.

We might label such a distribution an \textit{anti-normally
ordered} Unruh effect since  the vibrational
excitation from the ground to the excited state takes place by an
absorption of a photon and an emission of a phonon to the electronic state
$|e\rangle$ as depicted in  \Fig{fig1}{b},
i.e. by a term such as $\hat{a}_p \, \sigma_-$,  see Eq.(\ref{blue}). As discussed earlier,
such a detector responds to the anti-normally ordered moments
of the phonon field amplitude. Note that as $\kappa\to 0$ ($T\to 0$) in \Eq{14.5}
we get a \textit{finite} contribution to the probability for
excitation $P_m$. In the case of red sideband detuning, the usual
Unruh effect analogy in \Eq{14}, $P_m \to 0$ as $\kappa\to 0$.
This limit corresponds to a fixed trap frequency $\nu$ for which we get
no excitation as described above.

The above $\kappa\to 0$ limiting cases can also be understood as follows.
For a constant trap frequency
$\nu_p(t) = \nu_p$ the relevant integral to compute is
\be{14.6}
I_p(\infty,-\infty) = \int_{-\infty}^\infty \, dt \, e^{i \Delta t} \, e^{i\nu_p t} = 2\pi \delta (\Delta + \nu_p)
\ee
which arises from the $\hat{a}_p^\dagger\sigma_+$ term in $\hat{H}^{(m)}_I$ \Eq{5}
when acting on the motionally cooled ground state \Eq{6}.
For blue sideband detuning $\Delta = -|\Delta| < 0$ we obtain a non-zero contribution from the delta function.
In the usual ion trap excitation schemes for quantum computing \cite{wineland}, the resonant
(rotating wave approximation) portion of the Hamiltonian \Eq{5}
gives rise to the anti-Jaynes-Cummings type interaction \Eq{blue}, which permits
transitions of the form $|n\rangle|g\rangle \to |n+1\rangle|e\rangle$. With our motionally cooled ground state \Eq{6}
with $n=0$ such transitions are possible from $|0\rangle|g\rangle \to |1\rangle|e\rangle$. Thus
$\hat{a}_p^\dagger\sigma_+$ represents a resonant contribution to the Hamiltonian.

For red sideband detuning, $\Delta > 0$ we do not obtain a contribution from the delta function.
Here the resonant portion of the Hamiltonian \Eq{5}
gives rise to the usual Jaynes-Cummings type interaction Eq.(\ref{red}),  which permits
transitions of the form $|n\rangle|g\rangle \to |n-1\rangle|e\rangle$. With an initial motional ground state \Eq{6}
with $n=0$ such transitions are not possible. However, in writing down \Eq{red} 
we have dropped the non-resonant anti-normally ordered terms. It is the
$\hat{a}_p^\dagger\sigma_+$ non-RWA term in $\hat{H}^{(m)}_I$ \Eq{5}  that can produce transitions
from the motionally cooled ground state \Eq{6} and is responsible for the Unruh-like behavior of the probability $P_m$
for excitation out of this ground state.

We can also give an Unruh analogy interpretation
of the zero contribution to the delta function $\delta (\Delta + \nu_p)$ in \Eq{14.6}.
The case of red sideband detuning $\Delta > 0$,  a constant trap frequency is analogous to
an inertial observer moving with constant velocity in Minkowski space, giving rise to a constant Lorentz
transformation as discussed above. For a constant velocity observer, an inertial detector would have
a response function proportional to $\delta(E+E_0)$ where $E \; (\textrm{analogously}\; \Delta)$ is the energy of the detected particle
and $E_0>0 \; (\textrm{analogously}\; \nu_p)$ represents the atomic transition frequency producing the particle to be detected \cite{BD}.
Since the detector only responds to positive energies $E>0$, the contribution from the delta function $\delta(E+E_0)$ is zero,
which simply states that an inertial detector moving through the Minkowski vacuum would detect no particle production.

\section{Finite Chirp, General Expression}
For a finite chirp between times $t_0$ and $T$ we can develop a general
expression for $P_m(T,t_0)$ in terms of the incomplete gamma
functions $\gamma(\mu,x) = \int_0^x dt\; t^{\mu-1}
e^{-t}$ and $\Gamma(\mu,x) = \int_x^\infty dt\;
t^{\mu-1} e^{-t}$ such that $\gamma(\mu,x) +
\Gamma(\mu,x) = \Gamma(\mu)$. Let us write
$\int_{y_0}^{y_T} dy = \int_{0}^{\infty} dy - \int_{0}^{y_0} dy -
\int_{y_T}^{\infty} dy$
for general finite duration limits in \Eq{Ipy}
with the definitions $y_0 = e^{\kappa t_0}$ and
$y_T = e^{\kappa T}$.
By scaling the integration
variable to $y = y_0 x$ in the second integral on the right hand
side and $y = y_T x$ in the third integral on the right hand side
one can easily show that
\be{20} I_p(T,t_0) = \frac{1}{\kappa } \,
\displaystyle \frac{\Gamma(i a)  \left( e^{ i \frac{\pi}{2}  }\right)^{i a} }{a^{i a}}
\Big( 1 -\gamma'(i a,-i b y_0) - \Gamma'(i a, - i b y_T) \Big),
\ee
with the same definitions  of $a$ and $b$ used in \Eq{Ipy}, and where we
have defined the normalized gamma functions
$\gamma'(\mu,x) = \gamma(\mu,x)/\Gamma(\mu)$
and $\Gamma'(\mu,x) = \Gamma(\mu,x)/\Gamma(\mu)$ such that
$\gamma'(\mu,x) + \Gamma'(\mu,x) = 1$. Thus we
obtain
\be{21}
P_m(T,t_0) = \frac{\Omega_0^2\eta^2}{\kappa\Delta} \;
\frac{2\pi}{e^{\hbar\Delta/k_B T}-1} \;
\sum_{p=1}^N  \left| 1 -
\gamma'\left(i\frac{\Delta}{\kappa},-i \frac{\nu_p}{\kappa}
e^{\kappa t_0}\right)
   - \Gamma'\left(i\frac{\Delta}{\kappa}, - i \frac{\nu_p}{\kappa} e^{\kappa T}\right)\right|^2
\frac{|b_1^{(p)}|^2}{\sqrt{\mu_p}}.
\ee

The previous expression for the total excitation probability in \Eq{14}
corresponds to $P_m(\infty,-\infty)$ using \Eq{21} above which formally
corresponds to sweeping the trap frequency from an initial zero value to
an infinite final value. Considering a more realistic situation applicable
to experiments, let us consider $t_0 = 0$ and a finite trap expansion time $T$,
which corresponds to sweeping the trap frequency from $\nu \to \nu e^{\kappa T}$.
Considering \Eq{21} as a function of the detuning $\Delta$ with parameters
$\nu_p/\kappa$ and $\kappa T$ we can recover \Eq{14} under the following conditions
\be{22}
\frac{\nu_p}{\kappa} \ll 1, \qquad \frac{\nu_p}{\kappa}  e^{\kappa T} \gg 1 \qquad \forall\, p = (1,2,\ldots, N)
\ee
which makes the incomplete gamma functions small compared to unity.
As an example, taking $\nu_p/\kappa = 0.01$ and
$(\nu_p/\kappa) \exp(\kappa T) = 100$ requires that $\kappa T > 9.2$

\begin{figure}[h]
\includegraphics[width=4in,height=3in]{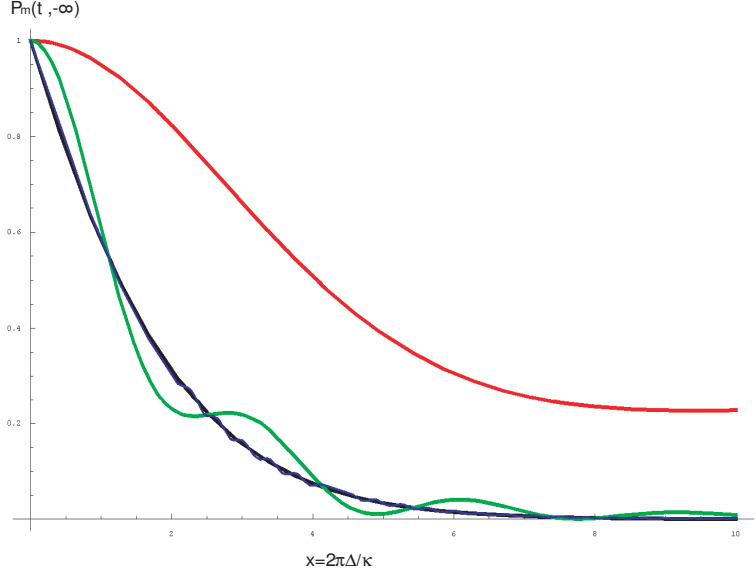}
\caption{Finite duration chirp: Plot of \Eq{21} for $p=1$ and
$\Omega_0\eta/\kappa = 1$, $\nu_p/\kappa=1$ with $x= 2\pi\Delta/\kappa$ as the
abscissa and $P_m(t,-\infty)$ as the ordinate for various values
of $y_t=e^{\kappa T} = \nu_1(T)/\nu$; red curve $y_T = 1$, green
curve $y_T = 10$, blue curve $y_T = 100$. The smooth black curve is the pure
Unruh case $P_m(\infty,-\infty)$ given by \Eq{14}. } \label{fig3}
\end{figure}

Note that \Eq{21} approaches $P_m(\infty,-\infty)$ of \Eq{14}
as $t_0\to -\infty$ and $T\to\infty$ in which the incomplete gamma functions approach zero.
The main point for experimental purposes is that we only need to take finite
limits of the order $\nu_p/\kappa \sim 10^{-2}$ and
$(\nu_p/\kappa) \exp(\kappa T) \sim 10^{2}$ to approximate the full Unruh case,
$P_m(\infty,-\infty)$ as shown in \Fig{fig3}.

\section{Discussion and conclusion.}
As we have shown, the experimental signature of the exponential modulation of the trap frequency is
the Planck-like form for the excitation probability for the two level electronic system in each ion.
In such experiments it is the ratio of the excitation probability on the red ($\Delta=\nu$) and blue ($|\Delta|=\nu$)sidebands that is determined:
\begin{equation}
R_e=\frac{P_m^R}{P_m^B}
\end{equation}
as this number is independent of the Rabi frequency, the Lamb-Dicke parameter, and the time of interaction between
the vibrational and electronic degrees of freedom\cite{Monroe95,Turchette2000}.  Let us consider the case of a
single ion with trap frequency $\nu$.  Using Eqs.(\ref{14},\ref{14.5}) we see that
\begin{equation}
R=\frac{1-e^{-2\pi\nu/\kappa}}{e^{2\pi\nu/\kappa}-1}
\end{equation}
In a typical experiment one can detect $R$ values as low as $0.05$ with about $20$\% error.  This
implies that $\nu/\kappa\approx 0.5$. At secular frequencies of $\nu/2\pi=0.1\ -\ 1$~MHz, we need
a modulation frequency of the order of a few hundred Khz to MHz; a not particularly difficult requirement for fast electronics.

The key issue however is the absolute size of the excitation probabilities at the red and blue sideband.
This is determined by the prefactor $(\Omega_0\eta)^2/(\kappa\nu)$. Defining
\begin{equation}
z=2\pi\nu/\kappa
\end{equation}
the equations for the excitation probability at the red and blue sidebands are
\begin{eqnarray}
P^R & = & \left (\frac{\Omega_0\eta}{\nu}\right )^2\frac{z}{e^z-1}\\
P^B &= & \left (\frac{\Omega_0\eta}{\nu}\right )^2\frac{z}{1-e^{-z}}
\end{eqnarray}
As we expect $z$ to be of the order of unity, we require that the secular frequency is within one order of magnitude
of the effective Rabi frequency.  This corresponds to a rather weakly bound ion, but should be achievable if stimulated
Raman transitions are used to couple the two level system. For example the relevant transition in $\mbox{}9$Be$\mbox{}^+$
can have an effective Rabi frequency of the order of $\Omega_0/2\pi=500$kHz\cite{Sackett2000}. If we use the centre of mass mode with
secular frequency of $\nu/2\pi=200$kHz, and a Lamb-Dicke parameter of $\eta=0.2$, the prefactor is $0.25$.
At a more conservative estimate of $\eta=0.05$, well within the Lamb-Dicke regime, the prefactor drops to a value
of $0.015$. These numbers are encouraging enough to suggest the plausibility of observing an analogous Unruh-like
effect in today's linear ion traps.

\acknowledgements
The authors would like to thank W. Hensinger and P. Milonni for helpful discussions.




\begin{thebibliography}{99}
\bibitem{DJ} D.F.V. James, Appl. Phys. B \textbf{66}, 181 (1998).
Note: Eq.(40) of this referecne is for a dipole transition, but
can also be used in our context as a quadrapole transition driven by
a Raman process, which modifies the definition of $\Omega_0$.
\bibitem{note1} \Eq{PmG} for the total excitation probability
stems from the  expression
$P_m(T,t_0) = \sum_k |A^{(m)}_k|^2$ where the amplitude for
a transition from the vacuum to any arbitrary exicted state $|\phi_k\rangle$
is given by the first order perturbation theory
$A^{(m)}_k = (-i/\hbar) \langle \phi_k|\int_{t_0}^T dt \hat{H}^{(m)}_I(t) | \mbf{0}\rangle$,
and use has been made of the completeness relation $\sum_k |\phi_k\rangle \langle \phi_k| = 1$.
\end{thebibliography}
\end{document}